\documentclass[aps,prl,superscriptaddress,showpacs,twocolumn]{revtex4-2}
\usepackage{bm}
\usepackage{graphicx}
\usepackage{color}
\usepackage{amsfonts}
\usepackage{amsmath}
\usepackage{amssymb,amsthm}
\usepackage{mathtools}
\usepackage{url}
\usepackage{hyperref}
\usepackage{booktabs}
\usepackage{multirow}
\usepackage{comment}

\newcommand{\ad}{\operatorname{ad}}

\definecolor{dgreen}{rgb}{0,0.5,0}

\begin{document}

% --------------------  TITLE  --------------------

\title{Central Charge in Quantum Optics}

% ------------  AUTHORS AND AFFILIATIONS ----------

\author{Daniel Burgarth}
%\email{daniel.burgarth@fau.de}
%\orcid{0000-0003-4063-1264}
\affiliation{Department Physik, Friedrich-Alexander-Universit\"at Erlangen-N\"urnberg, Staudtstra\ss e 7, 91058 Erlangen, Germany}
\author{Paolo Facchi}
%\email{paolo.facchi@ba.infn.it}
%\orcid{0000-0001-9152-6515}
\affiliation{Dipartimento di Fisica, Universit\`a di Bari, I-70126 Bari, Italy}
\affiliation{INFN, Sezione di Bari, I-70126 Bari, Italy}
\author{Hiromichi Nakazato}
%\email{hiromici@waseda.jp}
%\orcid{0000-0002-5257-7309}
\affiliation{Department of Physics, Waseda University, Tokyo 169-8555, Japan}
\author{Saverio Pascazio}
%\email{saverio.pascazio@ba.infn.it}
%\orcid{0000-0002-7214-5685}
\affiliation{Dipartimento di Fisica, Universit\`a di Bari, I-70126 Bari, Italy}
\affiliation{INFN, Sezione di Bari, I-70126 Bari, Italy}
\author{Kazuya Yuasa}
%\email{yuasa@waseda.jp}
%\orcid{0000-0001-5314-2780}
\affiliation{Department of Physics, Waseda University, Tokyo 169-8555, Japan}

%\date{\today}
%\date[]{September 10, 2024}

% --------------------  ABSTRACT  --------------------

\begin{abstract}
The product of two unitaries can normally be expressed as a single exponential through the famous Baker--Campbell--Hausdorff formula. 
We present here a counterexample in quantum optics, by showing that an expression in terms of a single exponential is possible only at the expense of the introduction of a new element (a central extension of the algebra), implying that there will be unitaries, generated by a sequence of gates, that cannot be generated by any time-independent quadratic Hamiltonian. A quantum-optical experiment is proposed that brings to light this phenomenon.
\end{abstract}
\maketitle

\textit{Introduction.---}%
Given an ensemble of (time-independent) Hamiltonians, consider the evolution obtained by applying them in a given temporal sequence. There are many examples of this sort in quantum information and applications, where one applies quantum gates or performs quantum operations~\cite{NC}, in evolutions and approximations involving the Trotter product formula~\cite{Trotter,BFGP}, and in topics related to the quantum Zeno effect~\cite{MS,FP}, and quantum control~\cite{QControl}.

It is often useful (and desirable) to find one (time-independent) Hamiltonian that generates the same evolution. Such a Hamiltonian normally belongs to the algebra of the initial set of operators (technically called the dynamical algebra of the system, obtained by taking linear combinations of commutators~\cite{QControl}). As we will discuss in this Letter, this is not true for optical gates generated by quadratic operators, describing the evolution of Gaussian states. In such a case, there will be unitaries, obtained by applying a sequence of gates, that cannot be generated by any time-independent quadratic Hamiltonian.

We will consider an explicit example that involves squeezing and phase shift. By using these two optical elements, we will obtain a transformation that cannot be obtained by applying only one optical element. Interestingly, we will see that this reflects a divergence of the Baker--Campbell--Hausdorff (BCH) series~\cite{ref:Hall-BCH,BCH2}, and is technically related to the appearance of a central extension of the operator algebra~\cite{centralc}, a phenomenon that we will thoroughly analyze in the following. We observe that such a phenomenon is familiar in quantum field theory~\cite{Weinberg1} and condensed-matter physics, but not in the context of quantum optics and quantum information.

Central extensions and central charges are physically interesting for several reasons. They arise in the study of symmetry groups and their representations, and can modify the algebra of the generators of symmetries and conservation laws, providing a deeper understanding of the underlying structure of physical theories.

In high-energy physics, they play a significant role in the study of conformal symmetries and conservation laws \cite{BB02,HH17}, as well as anomalies in quantum field theories \cite{Bertlmann00},  
that arise when classical symmetries are not preserved at the quantum level: in this context, they are crucial for maintaining the consistency of the theory. They are very relevant in grand unified theories \cite{BH10} and string theory \cite{EG19}.
In condensed-matter physics, central extensions are very important in the context of the topological phases of matter \cite{W17}, in the study of topological insulators \cite{P17}, and in the quantum Hall effect \cite{KS22}.
They are also closely related to the concept of fractionalization and the emergence of anyonic excitations \cite{K06,S08}, and play a prominent role in the formulation of the fractional quantum Hall effect, where the Laughlin wave function can be understood in terms of a Chern--Simons theory with a central extension \cite{FZ91,CD93}. 
We propose here a practical example in terms of a simple quantum-optical setting and discuss a possible experiment that allows a direct observation of this phenomenon.

\textit{Applicability of the BCH formula.---}%
The product of two unitaries is unitary and thus it is expected to be expressed as a single exponential,
\begin{equation}
e^Ae^B=e^C,
\label{eq:eAeBeC}
\end{equation}
where $A$, $B$, and $C$ are anti-Hermitian operators.
In the usual naive formulation of the BCH formula, one writes $C$ as a series expansion of nested commutators,
\begin{equation}
C=A+B+{1\over2}[A,B]+{1\over12}[A,[A,B]]+{1\over12}[[A,B],B]+\cdots.
\label{eq:Cexpanded}
\end{equation}
An important observation, in the present context, is that, if the commutators close, $C$ can be constructed in terms of the elements of the (closed) algebra. This is at the basis of the naive expectation that by taking commutators, as in the above BCH expansion, one does not ``leave'' the algebra. This is what will make the appearance of a central charge  counterintuitive in the following.

The BCH formula is derived under the assumption of ``small'' exponents. The convergence of the series~(\ref{eq:Cexpanded}) has been a subject of discussion (see, e.g., Ref.~\cite{ref:Blanes-BCH}), and there is no general treatment valid for ``large'' exponents, except for some simple solvable cases (a classical example being the case of $[A,B]$ commuting with both $A$ and $B$). It would therefore be desirable to have a nontrivial case where the exponent $C$ can be obtained analytically.
We stress that the group is connected: $e^A e^B$ is continuously connected to the identity, e.g.\ by the continuous path $U(s)= e^{sA} e^{sB}$, with $s \in [0,1]$, $U(0)=1$, and $U(1)= e^A e^B$.

In this Letter, we will derive an analytical form of $C$ in a particular, yet interesting physical case, and endeavor to shed light on the above issue. Our purpose is threefold.
First, we provide an illustrative example in quantum optics where the exponent $C$ requires an extension of the symplectic algebra $\mathrm{sp}(2,\mathbb{R})$. The parameters that appear in our analysis characterize the ``size'' of operators $A$ and $B$, that need not be small. 
Second, we examine the behavior of $C$, especially at large values of the parameters, and show that $C$ exhibits a bifurcation at a certain critical value of a parameter, which results in the introduction of a \emph{new} element that is \emph{not} included in the original algebra. 
A naive extension (or a wrong choice of the branch) would yield a diverging $C$ at a ``horizon'' beyond which no single exponent exists without the extension of the algebra.    
We will see that such a bifurcation usually only appears when the algebra is not compact. 
Finally, we discuss these results on the basis of an interferometric quantum-optical example.

\textit{Case study.---}%
Consider the quadratic quantum-optical Hamiltonians, representing phase shift~$h_H$ and squeezings~$h_\pm$,
\begin{equation}
h_H={1\over4}(\hat a^\dagger\hat a+\hat a\hat a^\dagger),\;h_{+}={1\over4}(\hat a^\dagger{}^2+\hat a^2),\;h_{-}={1\over4i}(\hat a^\dagger{}^2-\hat a^2),
\label{eq:metaplectic}
\end{equation}
where $\hat a$ and $\hat a^\dagger$ are the annihilation and creation operators, satisfying $[\hat a,\hat a^\dagger]=1$.
They form a closed algebra under the  commutation relations
\begin{equation}
[h_+,h_H]=-ih_-,\ \ [h_H,h_-]=-ih_+,\ \ [h_-,h_+]=ih_H.
\label{eq:halgebra}
\end{equation}
Notice that the operators $h_H$ and $h_\pm$ are Hermitian.
The one-parameter ($t$) actions of these Hamiltonians on $\hat x={1\over\sqrt2}(\hat a+\hat a^\dagger)$ and $\hat p={1\over\sqrt{2}\,i}(\hat a-\hat a^\dagger)$ are represented by the elements of the symplectic group $\mathrm{Sp}(2,\mathbb{R})$ (see, e.g., Ref.~\cite{Adesso}),
\begin{align}
e^{i t h_-}\begin{pmatrix}\hat x\\\hat p\end{pmatrix}e^{-i t h_-}&=e^{-t\sigma_3/2}\begin{pmatrix}\hat x\\\hat p\end{pmatrix},\\
e^{i t h_H}\begin{pmatrix}\hat x\\\hat p\end{pmatrix}e^{-i t h_H}&=e^{it\sigma_2/2}\begin{pmatrix}\hat x\\\hat p\end{pmatrix}.
\end{align}
The generators $\{-\sigma_1/2,\,i\sigma_2/2,\,-\sigma_3/2\}$ with $\sigma_i$ ($i=1,2,3$) the Pauli matrices, corresponding to $\{-i h_+,\,-i h_H,\,-i h_-\}$, are the elements of the algebra $\mathrm{sp}(2,\mathbb{R})$.
Clearly, these generators satisfy the same commutation relations as~(\ref{eq:halgebra}), but without $i$ on the right-hand sides.

The Lie algebra generated by the quantum-optical operators~(\ref{eq:metaplectic}) is an infinite-dimensional representation of the symplectic algebra $\mathrm{sp}(2,\mathbb{R})$, called metaplectic representation in the mathematical literature~\cite{Folland,Segal}. 
The unitaries generated by those Hamiltonians form in fact a covering group of the symplectic group [with a mechanism analogous to $\mathrm{SU}(2)$ and $\mathrm{SO}(3)$]. For a physically flavored discussion of this covering group, see e.g.~Ref.~\cite{Woit}.

Observe that the product of a particular $2\pi$ phase shift and a squeezing operator, which is surely an element of $\mathrm{Sp}(2,\mathbb{R})$, cannot be expressed as a single exponential of $\mathrm{sp}(2,\mathbb{R})$, because  
\begin{equation}
e^{-t\sigma_3/2}e^{\pi i\sigma_2}=\begin{pmatrix}-e^{-t/2}&0\\0&-e^{t/2}\end{pmatrix}=e^{-t\sigma_3/2+\pi i\hat1}.
\label{eq:centralext}
\end{equation}     
The extra identity operator $\hat1$ multiplied by $\pi i$, which is not an element of $\mathrm{sp}(2,\mathbb{R})$ and commutes with any of its elements, is necessary to express the product: this clearly shows a breakdown of the BCH formula. Technically, this signifies the appearance of a central extension of the operator algebra~\cite{centralc}, a phenomenon that we will analyze in the following. We observe that such a phenomenon is familiar in quantum field theory~\cite{Weinberg1} and condensed-matter physics, but not in the present quantum-optical context.

In order to gain additional insight into this issue, we look at a specific example, and endeavor to find an analytic expression of the exponent $C=ih$ for the product~(\ref{eq:eAeBeC}) of two unitaries  with $A=i\omega h_H$ and $B=i\eta h_-$,
\begin{equation}
e^{i\omega h_H}e^{i\eta h_-}=e^{ih},
\label{eq:h}
\end{equation}
where $\omega$ and $\eta$ are real parameters and $h$ is a Hermitian operator.  
By parametrizing the exponent as 
\begin{align}
h
&=a\cosh\xi\sin{\omega\over2} \, h_+ +a\cosh\xi\cos{\omega\over2} \, h_- +a\sinh\xi \, h_H\nonumber \\ 
&=\alpha(\omega,\eta)\, h_+ + \beta(\omega,\eta)\, h_- + \gamma(\omega,\eta)\, h_H,
\label{eq:h1}
\end{align}
we obtain the parameters $a$ and $\xi$ as the solutions of the equations
\begin{align}
\sinh{a\over2}\cosh\xi&=\sinh{\eta\over2},\label{eq:s1}\\
\sinh{a\over2}\sinh\xi&=\cosh{\eta\over2}\sin{\omega\over2},\label{eq:s2}\\
\cosh{a\over2}&=\cosh{\eta\over2}\cos{\omega\over2}.\label{eq:s3}
\end{align}
Notice that the parameters $a$ and $\xi$ are determined through~(\ref{eq:s1})--(\ref{eq:s3}) as functions of $\omega$ and $\eta$ and are in general complex-valued. Clearly, one must make sure that the solution, obtained by inverting the trigonometric functions in~(\ref{eq:s1})--(\ref{eq:s3}), yields a unitary operator in~(\ref{eq:h}).

\begin{figure}[t]
\includegraphics[width=6cm]{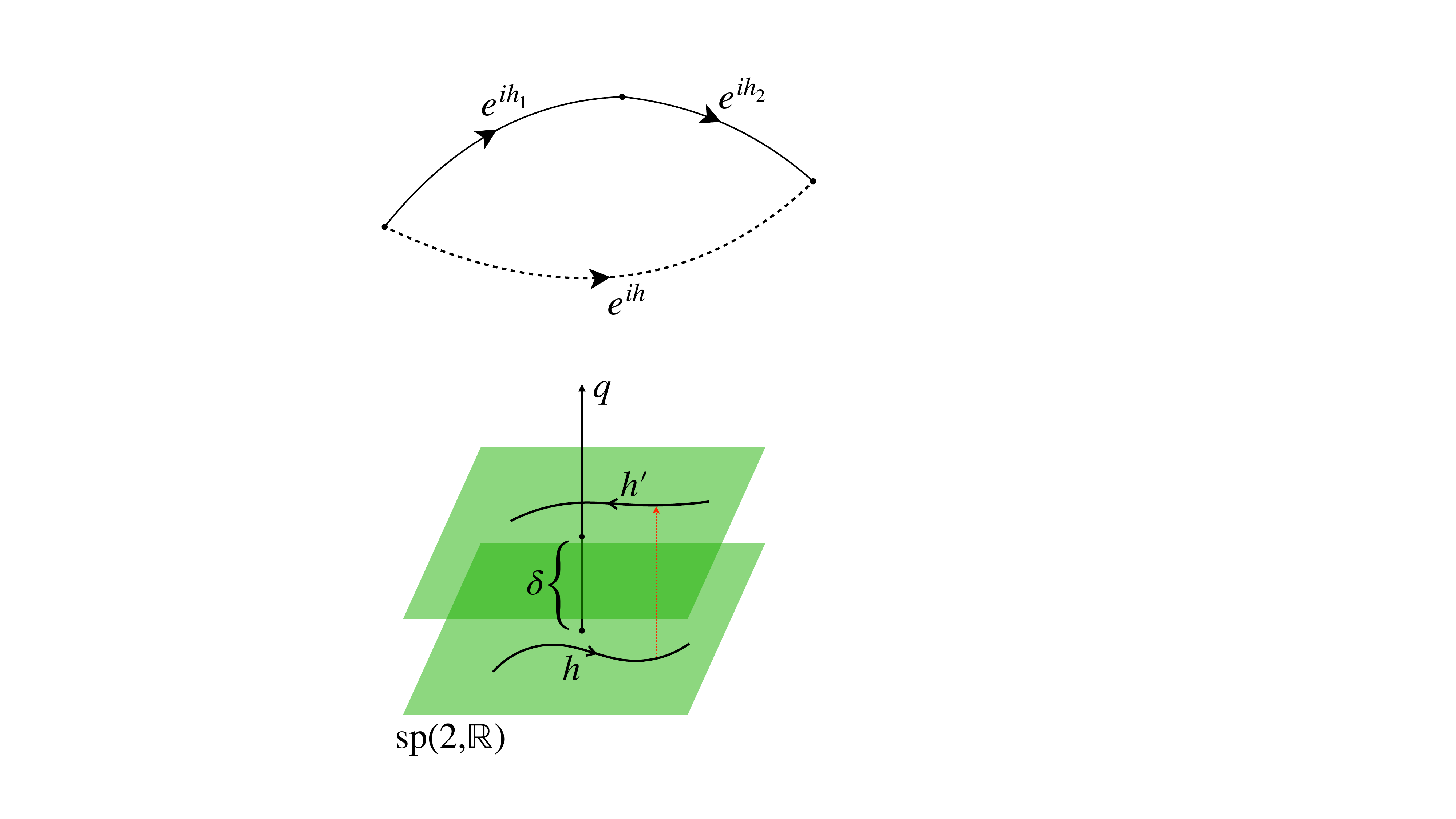}
\caption{Can the product of two unitaries $e^{ih_1}e^{ih_2}$ always be reproduced by a single unitary $e^{ih}$ in linear optics?}\label{fig:h1h2h}
\end{figure}

Summarizing, given the parameters $\omega$ and $\eta$, which determine the two unitaries on the left-hand side of~(\ref{eq:h}) and characterize the ``size'' of their exponents, their successive action can be reproduced by a unitary evolution generated by the Hamiltonian $h$, provided we can determine the (finite) coefficients in~(\ref{eq:h1}).
A schematic picture of the problem is shown in Fig.~\ref{fig:h1h2h}\@.

\begin{figure}[b]
\includegraphics[width=8 cm]{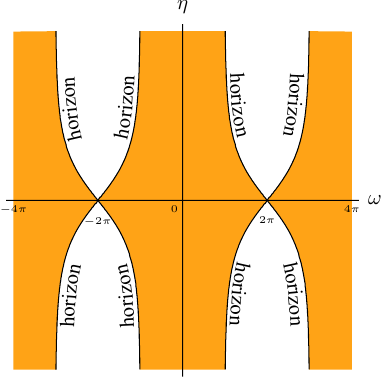}
\caption{Horizons in the $(\omega,\eta)$ plane. Along these critical curves, defined by~(\ref{eq:criticalhorizon}), the Hamiltonian $h$ in~(\ref{eq:h1}) diverges.
Beyond the horizons it becomes impossible to express the Hamiltonian $h$ as a linear combination of the elements of the algebra~(\ref{eq:halgebra}), and a \emph{central extension} is needed.
}
\label{fig:horizon}
\end{figure}

\textit{Solution and validity range: The central charge.---}%
The full solution of the problem is involved, and is detailed in the Supplemental Material~\cite{SM}. We summarize here its most salient features.

It is straightforward to explicitly obtain $h$ for small values of $\omega$ and $\eta$. On the other hand, when the parameters $\omega$ and $\eta$ become larger, the problem becomes much more involved. One observes the presence of \emph{critical curves} in the $(\omega,\eta)$ plane, that are solutions of the equation 
\begin{equation}
\cosh{\eta\over2}\cos{\omega\over2} = -1.
\label{eq:criticalhorizon}
\end{equation}
These curves can be viewed as \emph{horizons}, in the sense that the coefficients appearing in~(\ref{eq:h1}) diverge at these curves. 
This is shown in Fig.~\ref{fig:horizon}\@. Beyond the horizons it becomes impossible to express $h$ as a linear combination of the elements of the algebra~(\ref{eq:halgebra}), and a \emph{central extension} is needed to write $h$. 
The phenomenon is similar to that described in~(\ref{eq:centralext}). The continuation of the coefficients to and through the horizon in Fig.~\ref{fig:horizon} is studied in the Supplemental Material~\cite{SM}. The Hamiltonian beyond the horizons reads
\begin{equation}
h = \alpha'(\omega,\eta)h_+ + \beta'(\omega,\eta) h_- + \gamma'(\omega,\eta) h_H + \delta \, q,
\label{eq:h2}
\end{equation}
where the novel coefficients are primed and $\delta=\pi/2$ is a constant. Notice the presence of the additional operator
\begin{equation}
q = 2 - (-1)^{\hat n} = 1+2P_\mathrm{odd},
\label{eq:q}
\end{equation}
with $\hat n = \hat a^\dagger \hat a$ the number operator.
Observe that $q$ is related to the parity operator, with $P_\mathrm{odd}$ the projection operator onto the odd-number states. The operator $q$ does not belong to the original algebra~(\ref{eq:halgebra}), and commutes with all its elements. It is needed in order to properly express the Hamiltonian beyond the horizons in Fig.~\ref{fig:horizon}\@.
For instance, for $\omega=2\pi$ and $\eta=t$, we get
\begin{equation}
e^{\pi i\hat{a}^\dag\hat{a}}e^{t(\hat{a}^\dag{}^2-\hat{a}^2)/4}=e^{t(\hat{a}^\dag{}^2-\hat{a}^2)/4+\pi iP_\mathrm{odd}}.
\end{equation}
This is the counterpart of~(\ref{eq:centralext}).
Note that $P_\mathrm{odd}$ is not quadratic in $\hat{a}$ or $\hat{a}^\dag$.
\begin{figure}[b]
\includegraphics[width=7 cm]{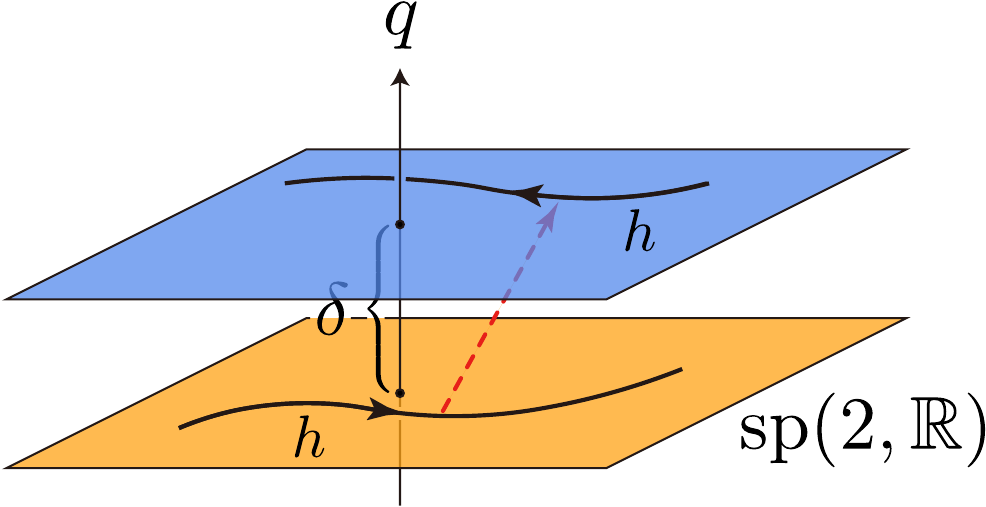}
\caption{Manifold spanned by the symplectic algebra $\mathrm{sp}(2,\mathbb{R})$ in~(\ref{eq:halgebra}) and the additional ``direction'' $q$.
By approaching the horizon, the expression~(\ref{eq:h1}) of the Hamiltonian $h$ in terms of the coefficients $\alpha$, $\beta$, and $\gamma$ diverges, ceases to be valid, and must be replaced by~(\ref{eq:h2}), which includes the central charge $q$. The upper manifold in the figure contains the regions beyond the horizons in Fig.~\ref{fig:horizon}. In the upper manifold, the product of two unitaries $e^{i\omega h_H}e^{i\eta h_-}$ cannot be expressed as a single exponential of a linear combination of operators in the original algebra $\mathrm{sp}(2,\mathbb{R})$.}
\label{fig:horizoninfty}
\end{figure}

One may get an alternative pictorial representation of the phenomenon: see Fig.~\ref{fig:horizoninfty}, in which we represent the  manifold spanned by the (three-dimensional) symplectic algebra $\mathrm{sp}(2,\mathbb{R})$ in~(\ref{eq:halgebra}) and an additional ``direction'' along $q$. Let us follow the evolution of a given trajectory of the Hamiltonian $h$ in~(\ref{eq:h1}), parametrized by $\omega$ and $\eta$. 
By approaching the horizon (at \emph{finite} values of $\omega$ and $\eta$), the Hamiltonian $h$ and its coefficients $\alpha$, $\beta$, and $\gamma$ diverge, the expression~(\ref{eq:h1}) ceases to be valid, and must be replaced by~(\ref{eq:h2}), which includes the additional operator (central charge) $q$. This is the upper manifold displayed in Fig.~\ref{fig:horizoninfty}, and contains the regions beyond the horizons in Fig.~\ref{fig:horizon}.

Actually, one can confirm that there are two solutions of (\ref{eq:s1})--(\ref{eq:s3}) coexisting in the region $-1< \cosh{\eta\over2}\cos{\omega\over2} < 1$~\cite{SM}.
One of them is just a smooth extension from the region $\cosh{\eta\over2}\cos{\omega\over2} > 1$ with smaller $\omega$, connected with $\openone$ at the origin $(\omega,\eta)=(0,0)$, and faces a singularity at the horizon $\cosh{\eta\over2}\cos{\omega\over2} = -1$, as already pointed out in~(\ref{eq:criticalhorizon}).
The other solution, on the other hand, remains finite at the horizon $\cosh{\eta\over2}\cos{\omega\over2} = -1$, but faces a singularity at the curve $\cosh{\eta\over2}\cos{\omega\over2} = 1$, and is unable to reach the origin $(\omega,\eta)=(0,0)$.
We stress that the point $\omega=\pi$ is unique, in the sense that the exponents $h$ of the two solutions coincide only at $\omega=\pi$, where we can choose an alternative branch keeping the exponent $h$ continuous.

\textit{Experimental proposal.---}%
The phenomenon we described is interesting from an abstract point of view, but it would be very interesting if one could conceive an experiment able to bring it to light. 
One possible option is the following. 
Consider an interferometric experiment, in which one branch wave interacts with a squeezer $h_-$ and then a phase shifter $h_H$, as on the left-hand side of~(\ref{eq:h}), while the other branch wave interacts with a Trotterized version of Hamiltonian $h$ in~(\ref{eq:h2}).

Consider the point $(\omega,\eta)=(2\pi-\epsilon,\eta)$ on the $(\omega,\eta)$ plane in Fig.~\ref{fig:horizon} with $\epsilon=0.1$ and $\eta = 0.2$, so that we are in the ``white'' region of parameters beyond the first horizon, close to the ``crossing point'' $(\omega,\eta) = (2\pi,0)$ on the horizontal axis. Incidentally, observe that the squeezing is quite small compared to available technology~\cite{squeezing}.
The exponential of the Hamiltonian $h$ in~(\ref{eq:h2}) without the parity operator $q$ can be Trotterized as
\begin{equation}
\left(e^{i \alpha'(\omega,\eta) h_+/N} e^{i \beta'(\omega,\eta) h_-/N} e^{i \gamma'(\omega,\eta) h_H/N} 
\right)^N ,
\label{eq:trotterreply}
\end{equation}
where $(\alpha', \beta', \gamma')$ are explicitly given by $(\alpha_2, \beta_2, \gamma_2)$ in~(60)--(61) of the SM [expressions valid beyond the first horizon, close to the point $(\omega, \eta) = (2\pi, 0)$].
We numerically computed the operator-norm difference between the Trotter product~(\ref{eq:trotterreply}) and 
$e^{-\pi iq/2}e^{i h}=  e^{-\pi iq/2}e^{i\omega h_H}e^{i\eta h_-}$
in the symplectic representation, to check the error of the Trotter approximation~(\ref{eq:trotterreply}). Observe that for small $\epsilon$ and $\eta$ the coefficients $\alpha'$, $\beta'$, and $\gamma'$ are small, and we expect the Trotterization to work efficiently.
We numerically found that for $\epsilon=0.1 $ and $\eta = 0.2$ the error of the Trotter approximation~(\ref{eq:trotterreply}) is $\simeq0.5 \%$ already for $N=1$ (!). It becomes $\simeq 0.25 \%$ for $N=2$ and $\simeq 0.1 \%$ for $N=5$, in agreement with the expectation that this difference should scale like $1/N$~\cite{Suzuki1985,Zagrebnov2024,BFGP}.

The experiment then consists in comparing the Trotter product~(\ref{eq:trotterreply}) (e.g.~for $N=1$) with $e^{i h}=e^{i\omega h_H}e^{i\eta h_-}$, in an interferometer.
This extracts the exponential $e^{\pi iq/2}$ of the central charge $q$.
As outlined after~(\ref{eq:q}), $q$  is essentially the parity operator.
Thus, depending on the parity of the input photon number $\hat{n}$, we observe different interference at the output of the interferometer.
This allows us to measure the central charge $q$.

\textit{Concluding remarks.---}%
We have considered a particular case of~(\ref{eq:eAeBeC}) with $A\propto i h_H$ and $B\propto ih_-$, as in~(\ref{eq:h}). However, since the operators $ih_\pm$ and $h_H$ satisfy the commutation relations of angular momentum and the strategy adopted in this Letter to derive $ih$ depends only on algebraic relations, our results and conclusions can be easily generalized to other cases, with different $A$ and $B$ proportional to different elements of the algebra, just by performing an appropriate ``rotation'' among $ih_\pm$ and $h_H$ together with the necessary change (or analytic continuation) of the parameters. In this sense, our outcomes are not limited to the particular case considered in~(\ref{eq:h}), but have a general character.

We also observe that by further increasing (or decreasing) $\omega$ (and/or $\eta$) in Fig.~\ref{fig:horizon}, one can explore other regions of the central extension. It would be interesting to analyze whether the presence of a number of horizons/manifolds entails physical consequences and how they could be experimentally verified.

Observe that the phenomenon we described does not occur if all the operators $ih_\pm$ and $h_H$ are Hermitian.
In such a case, they are essentially angular momenta and no central extension appears.
The singular behavior discussed in this Letter is solely due to the appearance of different signs in the commutation relations~(\ref{eq:halgebra}), reflecting the noncompactness of the algebra. 

Central extensions have long been known in physics. They provide a powerful framework for understanding some fundamental aspects of quantum field theories, symmetries, anomalies, and their applications in various branches of high-energy physics. They are also crucial in unveiling the features of the topological, symmetry-protected, and fractionalized phases of matter in condensed-matter physics. The beautiful discussion by Weinberg~\cite{Weinberg1} elucidates this mathematical phenomenon and its physical significance. 
We have shown here that the presence of central extensions can be brought to light in a simple quantum-optical setting, and is amenable to direct experimental verification.

\bigskip
\acknowledgments
This work is supported by the Top Global University Project from the Ministry of Education, Culture, Sports, Science and Technology (MEXT), Japan.
PF and SP acknowledge support from Istituto Nazionale di Fisica Nucleare (INFN) through the project `QUANTUM'.
PF acknowledges support from PNRR MUR project CN00000013 `Italian National Centre on HPC, Big Data and Quantum Computing' and from the Italian National Group of Mathematical Physics (GNFM-INdAM).
SP acknowledges support from PNRR MUR project PE0000023-NQSTI\@.
KY acknowledges support through JSPS KAKENHI Grant Numbers JP18K03470, JP18KK0073, and JP24K06904, from the Japan Society for the Promotion of Science (JSPS).

%\bibliography{central_ext.bib}
%\end{document}

%\end{document}

\newpage
\begin{widetext}
\appendix
\section*{Supplemental Material}
\subsection{Baker--Campbell--Hausdorff Formula}
The Baker--Campbell--Hausdorff (BCH) formula is a cornerstone in theoretical and mathematical physics~\cite{ref:Hall-BCH}.
One endeavors to express the product of two unitaries as a single exponential (unitary),
\begin{equation}
e^Ae^B=e^C.
%\label{eq:eAeBeC}
\end{equation}
In the usual naive formulation of the BCH formula, one writes down $C$ as a series expansion of nested commutators,
\begin{equation}
C=A+B+{1\over2}[A,B]+{1\over12}[A,[A,B]]+{1\over12}[[A,B],B]+\cdots.
\label{eq:Cexpanded_SM}
\end{equation}
From a rigorous perspective, the correct formula reads
\begin{equation}
C=A+\int_0^1dt\,g(e^{\ad_A}e^{{t\ad}_B})B,
\label{eq:Cexpandedrig}
\end{equation}
where $\ad_X\bullet=[X,{}\bullet{}]$ and 
\begin{equation}
g(z)={z\log z\over z-1}=1+{1\over2}(z-1)-{1\over6}(z-1)^2+{1\over12}(z-1)^3-\cdots.
\end{equation}
This function $g(z)$ is holomorphic in the disk $|z-1|<1$, and the BCH formula (\ref{eq:Cexpanded_SM}) is only valid for ``small'' exponents. 
There is no general treatments valid for ``large'' exponents, except for some simple solvable cases 
(a classical example being the case of $[A,B]$ commuting with both $A$ and $B$). 
The convergence of the series~(\ref{eq:Cexpanded_SM})--(\ref{eq:Cexpandedrig}) has been a subject of discussion (see, e.g., Ref.~\cite{ref:Blanes-BCH}).

\subsection{Derivation of the Single Hamiltonian}
We first remark that the Hamiltonians $h_\pm$ and $h_H$, introduced in~(3) of the main text, essentially satisfy the commutation relations for angular momenta.
In fact, $\{ih_+,ih_-,h_H\}$ do.
We therefore consider three operators $X\,(=ih_+)$, $Y\,(=ih_-)$, and $Z\,(=h_H)$, which satisfy the commutation relations for angular momenta
\begin{equation}
[X,Y]=iZ,\quad[Y,Z]=iX,\quad[Z,X]=iY.
\label{eq:AMalgebra}
\end{equation}
It is to be noticed that they are not necessarily Hermitian and can be infinite-dimensional.
Specifically, we consider 
\begin{equation}
X={i\over4}(\hat a^\dagger{}^2+\hat a^2)=-X^\dagger,\quad
Y={1\over4}(\hat a^\dagger{}^2-\hat a^2)=-Y^\dagger,\quad
Z={1\over4}(\hat a^\dagger\hat a+\hat a\hat a^\dagger)=Z^\dagger,
\label{eq:genrep}
\end{equation}
where operators $\hat a$ and $\hat a^\dagger$ are the usual bosonic annihilation and creation operators satisfying $[\hat a,\hat a^\dagger]=1$.
The goal is to find an analytic expression of the exponent $C$ for the following product of two unitaries
\begin{equation}
e^{i\omega tZ}e^{\eta t Y}=e^{C(t)},
\label{eq:defW}
\end{equation}
where $\omega$, $\eta$, and $t$ are real parameters.
In order to find the conditions that should be satisfied by $C$, take the derivative with respect to $t$ of both sides of~(\ref{eq:defW}).
The left-hand side gives
\begin{equation}
{d\over dt}(e^{i\omega tZ}e^{\eta t Y})
=[i\omega Z+\eta(X\sin\omega t+Y\cos\omega t)]e^{i\omega tZ}e^{\eta tY}
=[(\eta\bm e_1+i\omega\bm e_3)\cdot\bm\Sigma]e^{i\omega tZ}e^{\eta tY},
\end{equation}
where $\bm\Sigma\equiv(X,Y,Z)$, and we have introduced three unit vectors
\begin{equation}
\bm e_1=(\sin\omega t,\cos\omega t,0),\quad
\bm e_2=(-\cos\omega t,\sin\omega t,0),\quad
\bm e_3=(0,0,1).
\end{equation}
On the other hand, the right-hand side of~(\ref{eq:defW}) yields
\begin{equation}
{d\over dt}e^C=\left(
\dot C+{1\over2!}[C,\dot C]+{1\over3!}[C,[C,\dot C]]+\cdots
\right)e^C
=\left(
{e^{\ad_C}-1\over\ad_C}\dot C
\right)e^C. 
\end{equation}
Since the operators are closed with respect to the commutation relations, the operator $C$ is expected to be given as a linear combination of $\bm\Sigma$.
Introduce a parameter function $a(t)$ and a unit-vector function $\bm n(t)$ to express $C(t)$ as
\begin{equation}
C(t)=a(t)\bm n(t)\cdot\bm\Sigma+i\alpha(t)I,
\label{eq:parW}
\end{equation}
where the last term with a real parameter $\alpha(t)$ is proportional to an identity operator $I$, and turns out to play a crucial role in order that the operator $e^C$ covers all parameter regions of $e^{i\omega tZ}e^{\eta tY}$.
The derivative of $e^C$ reads
\begin{equation}
{d\over dt}e^C=\Bigl(
[\bm n\dot a+\dot{\bm n}\sinh a+i(\bm n\times\dot{\bm n})(\cosh a-1)]\cdot\bm\Sigma+i\dot\alpha I
\Bigr)\,e^C.
\label{eq:deW}
\end{equation}
That is, we have to find $a(t)$, $\bm n(t)$, and $\alpha(t)$ that satisfy
\begin{equation}
\bm n\dot a+\dot{\bm n}\sinh a+i(\bm n\times\dot{\bm n})(\cosh a-1)=\eta\bm e_1+i\omega\bm e_3,\quad
\bm n^2=1,\quad\dot\alpha=0,
\label{eq:condition}
\end{equation}
under the initial condition $e^{C(0)}=1$.
[We may replace the right-hand side of (\ref{eq:defW}) by $e^{C(t)}e^{-C(0)}$ for an arbitrary initial condition, which, however, is again a product of two factors and will not be considered in what follows.]
We understand that this would require the initial conditions $a(0)=0$ and $\alpha(0)=0$.
It is worth mentioning that even though the parameter $\alpha$ is constant with respect to $t$, this does not exclude the possibility that its (constant) value might depend on the parameter region and change abruptly at some points.

Since $C$ is expected to be anti-Hermitian, the third component of $\bm n$ is purely imaginary for a Hermitian $Z$. We introduce the following parametrization (valid for $\eta>\omega>0$, in the sense that, for small $t$, real-valued solutions would be available for $a$, $\xi$, and $\varphi$; this condition will however be relaxed afterwards)
\begin{equation}
\bm n=\cosh\xi\,(\bm e_1\sin\varphi+\bm e_2\cos\varphi)+i\bm e_3\sinh\xi.
\end{equation}
The above condition (\ref{eq:condition}) is explicitly written down as  
\begin{align}
\eta&=\dot a\cosh\xi\sin\varphi+\dot\xi[\sinh a\sinh\xi\sin\varphi-(\cosh a-1)\cos\varphi]+(\dot\varphi+\omega)\cosh\xi\,[\sinh a\cos\varphi-(\cosh a-1)\sinh\xi\sin\varphi],\nonumber\\
0&=\dot a\cosh\xi\cos\varphi+\dot\xi[\sinh a\sin\varphi\cos\varphi+(\cosh a-1)\sin\varphi]-(\dot\varphi+\omega)\cosh\xi\,[\sinh a\sin\varphi+(\cosh a-1)\sinh\xi\cos\varphi],\nonumber\\
\omega&=\dot a\sinh\xi+\dot\xi\sinh a\cosh\xi-(\dot\varphi+\omega)\cosh^2\xi\,(\cosh a-1).
\end{align}
A straightforward (though a bit lengthy) calculation yields
\begin{align}
\dot a&=\eta\cosh\xi\sin\varphi-\omega\sinh\xi,\nonumber\\
\dot\xi&={1\over2}
\left(
\coth{a\over2}\,(-\eta\sinh\xi\sin\varphi+\omega\cosh\xi)+\eta\cos\varphi
\right),\nonumber\\
\dot\varphi\cosh\xi&={1\over2}
\left(
\eta\coth{a\over2}\cos\varphi-\eta\sinh\xi\sin\varphi-\omega\cosh\xi
\right).
\label{eq:dots}
\end{align}

It is in general quite difficult to find integration factors that would make these equations completely integrable.
It is, however (interestingly) possible to show that the following quantities constitute a closed set with respect to differentiation
\begin{equation}
{d\over dt}
\begin{pmatrix}
\smallskip
\sinh{a\over2}\cosh\xi\cos\varphi\\
\smallskip
\sinh{a\over2}\cosh\xi\sin\varphi\\
\smallskip
\sinh{a\over2}\sinh\xi\\
\cosh{a\over2}
\end{pmatrix}=
\begin{pmatrix}
\smallskip
0&{\omega\over2}&{\eta\over2}&0\\
\smallskip
-{\omega\over2}&0&0&{\eta\over2}\\
\smallskip
{\eta\over2}&0&0&{\omega\over2}\\
0&{\eta\over2}&-{\omega\over2}&0
\end{pmatrix}
\begin{pmatrix}
\smallskip
\sinh{a\over2}\cosh\xi\cos\varphi\\
\smallskip
\sinh{a\over2}\cosh\xi\sin\varphi\\
\smallskip
\sinh{a\over2}\sinh\xi\\
\cosh{a\over2}
\end{pmatrix}.
\label{eq:diffeq}
\end{equation}
By observing that this $4\times4$ matrix, denoted as $M$, has the following $2\times2$ block form 
\begin{equation}
M=\begin{pmatrix}
\smallskip
{i\omega\over2}\sigma_2&{\eta\over2}\openone\\
{\eta\over2}\openone&{i\omega\over2}\sigma_2
\end{pmatrix},
\end{equation}
$\sigma_2$ being the second Pauli matrix, its exponential is easily evaluated to be
\begin{equation}
e^{Mt}
=\begin{pmatrix}
\smallskip
\cosh{\eta t\over2}\,e^{i{\omega t\over2}\sigma_2}&
\sinh{\eta t\over2}\,e^{i{\omega t\over2}\sigma_2}\\
\sinh{\eta t\over2}\,e^{i{\omega t\over2}\sigma_2}&
\cosh{\eta t\over2}\,e^{i{\omega t\over2}\sigma_2}
\end{pmatrix},
\end{equation}
where, of course, 
\begin{equation}
e^{i{\omega t\over2}\sigma_2}
=\begin{pmatrix}
\smallskip
\cos{\omega t\over2}&\sin{\omega t\over2}\\
-\sin{\omega t\over2}&\cos{\omega t\over2}
\end{pmatrix}.
\end{equation}
The initial condition $a(0)=0$ [and $\alpha(0)=0$] fixes the solution of (\ref{eq:diffeq}) uniquely and it is given by the fourth column of  $e^{Mt}$, i.e., $\alpha=0$ and
\begin{align}
\sinh{a\over2}\cosh\xi\cos\varphi&=\sinh{\eta t\over2}\sin{\omega t\over2},\nonumber\\
\sinh{a\over2}\cosh\xi\sin\varphi&=\sinh{\eta t\over2}\cos{\omega t\over2},\nonumber\\
\sinh{a\over2}\sinh\xi&=\cosh{\eta t\over2}\sin{\omega t\over2},\nonumber\\
\cosh{a\over2}&=\cosh{\eta t\over2}\cos{\omega t\over2}.
\end{align}
These are the exact expressions of the parameters $a$, $\xi$, and $\varphi$, which can be further reduced to 
\begingroup
\allowdisplaybreaks
\begin{align}
\varphi&={\pi\over2}-{\omega t\over2},\label{eq:sol0}\\
\sinh{a\over2}\cosh\xi&=\sinh{\eta t\over2},\label{eq:sol1}\\
\sinh{a\over2}\sinh\xi&=\cosh{\eta t\over2}\sin{\omega t\over2},\label{eq:sol2}\\
\cosh{a\over2}&=\cosh{\eta t\over2}\cos{\omega t\over2}.\label{eq:sol3}
%\label{eq:solution}
\end{align}
\endgroup

\subsection{Realization of the BCH Formula at Small $t$}
We first observe that the exponent $C$ defined as the logarithm of the product of two unitaries (\ref{eq:defW}) and parametrized as in (\ref{eq:parW}) is explicitly expressed as
\begin{equation}
C=a(\bm e_1\cosh\xi\sin\varphi+\bm e_2\cosh\xi\cos\varphi+i\bm e_3\sinh\xi)\cdot\bm\Sigma+i\alpha I
=Xa\cosh\xi\sin{\omega t\over2}+Ya\cosh\xi\cos{\omega t\over2}+iZa\sinh\xi+i\alpha I.
\label{eq:explicitW}
\end{equation}
The parameters are found (by paying due attention to higher-order terms necessary to evaluate square roots up to a prescribed order) to have the following small-$t$ expansions up to $O(t^3)$ ($\alpha=0$),
\begin{align}
a&=\sqrt{\eta^2-\omega^2}\,t
\left(
1+{\eta^2\omega^2\over48(\eta^2-\omega^2)}t^2
\right),\nonumber\\
\cosh\xi&={\eta\over\sqrt{\eta^2-\omega^2}}
\left(
1+{\omega^2(\eta^2-2\omega^2)\over48(\eta^2-\omega^2)}t^2
\right),\nonumber\\
\sinh\xi&={\omega\over\sqrt{\eta^2-\omega^2}}
\left(
1+{\eta^2(4\eta^2-5\omega^2)\over48(\eta^2-\omega^2)}t^2
\right).
\end{align}
Their combinations appearing in $C$ are thus expanded as
\begin{equation}
a\cosh\xi\sin{\omega t\over2}={\eta\omega\over2}t^2,\quad
a\cosh\xi\cos{\omega t\over2}=\eta t-{\eta\omega^2\over12}t^3,\quad
a\sinh\xi=\omega t+{\eta^2\omega\over12}t^3,
\label{eq:samllt}
\end{equation}
which implies that the exponent $C$ actually coincides with the form predicted by the Baker--Campbell--Hausdorff formula up to $O(t^3)$,
\begin{equation}
C=i\omega tZ+\eta tY+{1\over2}[i\omega tZ,\eta tY]+{1\over12}[i\omega tZ,[i\omega tZ,\eta tY]]+{1\over12}[[i\omega tZ,\eta tY],\eta tY].
\label{eq:BCH}
\end{equation}
Notice that the results (\ref{eq:samllt}) and (\ref{eq:BCH}) are independent of the assumption $\eta>\omega>0$, which has however been tacitly assumed in their derivation.

\subsection{Bifurcation and Central Charge}

In what follows, the parameter $t$ is set equal to 1.
Observe that the quantity $x=\cosh{\eta\over2}\cos{\omega\over2}$ first vanishes at $\omega=\pi$ when $\omega$ is increased from 0.
Quantities at $\omega=\pi$ shall henceforth be denoted with a subscript $0$, so that $a_0=\pm\pi i$, $\cosh{a_0\over2}=0$, $\sinh{a_0\over2}=\pm i$, and 
\begin{equation}
e^{C_0}=e^{\pi X\sinh{\eta\over2}+i\pi Z\cosh{\eta\over2}}.
\end{equation}
It is crucial to observe that if $a$ satisfies (\ref{eq:sol1})--(\ref{eq:sol3}) with the condition $a(0)\equiv a|_{\omega=\eta=0}=0$, then so does also the shifted quantity $a+2\pi i$
 [as can easily be seen in the differential equation (12)], with the initial condition shifted by $2\pi i$.
Therefore, we can write, for such a solution $a$ with $a(0)=0$,
\begin{equation}
e^{C}=e^{a\bm n\cdot\bm\Sigma}
=e^{(a+2\pi i)\bm n\cdot\bm\Sigma}e^{-2\pi i\bm n(0)\cdot\bm\Sigma},
\end{equation}
where $\bm\Sigma\equiv(X,Y,Z)$ formally satisfies the commutation relations among angular momenta.
This relation implies that $e^{2\pi i\bm n\cdot\bm\Sigma}$ is independent of vector $\bm n$, provided $\bm n^2=1$.
Furthermore, this quantity commutes with any operator in $\bm\Sigma$, i.e.,
\begin{equation}
e^{2\pi i\bm n\cdot\bm\Sigma}(\bm m\cdot\bm\Sigma)e^{-2\pi i\bm n\cdot\bm\Sigma}
=\bm m\cdot\bm\Sigma-[(\bm n\times\bm m)\cdot\bm\Sigma]\sin2\pi
+[(\bm n\times(\bm n\times\bm m))\cdot\bm\Sigma](1-\cos2\pi)
=\bm m\cdot\bm\Sigma,
\label{eq:2pi}
\end{equation}
for any vectors $\bm m$ and $\bm n$ with $\bm n^2=1$.
In this sense, it can be considered to be proportional to an identity operator \emph{in the space spanned by} $\bm\Sigma$.
Notice, however, that it satisfies the relations
\begin{equation}
e^{2\pi i\bm n\cdot\bm\Sigma}
\begin{pmatrix}
\smallskip
\hat a\\
\hat a^\dagger
\end{pmatrix}
e^{-2\pi i\bm n\cdot\bm\Sigma}
=-\begin{pmatrix}
\smallskip
\hat a\\
\hat a^\dagger
\end{pmatrix},\quad\forall\bm n,\;\bm n^2=1,
\end{equation} 
so that it anti-commutes with $\hat a$ and $\hat a^\dagger$ (that do not belong to the afore-mentioned algebra).
(Its square commutes with any operator in the total algebra, and is just the identity operator $e^{4\pi i\bm n\cdot\bm\Sigma}\propto1$.)

Now remember that at $\omega=\pi$, the value of $a$ is, say, $\pi i$, i.e.,
\begin{equation}
e^{C_0}=e^{i\pi\bm n_0\cdot\bm\Sigma}=e^{-i\pi\bm n_0\cdot\bm\Sigma}e^{2\pi i\bm n_0\cdot\bm\Sigma}.
\end{equation} 
Since it is trivial within the operator space spanned by $\bm\Sigma$, the last exponential factor can be interpreted as the exponential of an identity (times a constant), implying the appearance of an extra element in the exponent for $\omega\ge\pi$.
It is stressed that the exponent $C$ itself is continuous at $\omega=\pi$.
We have found explicitly 
\begin{equation}
e^{C}=e^{2\pi i\bm n_0\cdot\bm\Sigma}e^{\tilde aX\sinh\tilde\xi\sin{\omega\over2}
+\tilde aY\sinh\tilde\xi\cos{\omega\over2}+i\tilde aZ\cosh\tilde\xi},
\label{eq:h2BIS}
\end{equation}
for $\omega>\pi$ and $x>-1$, in terms of real parameters $\tilde a$ and $\tilde\xi$ defined according to $a=i\tilde a+2\pi i$ and $\xi=\tilde \xi-{\pi\over2}i$.

Since the quantity $e^{2\pi i\bm n_0\cdot\bm\Sigma}$ can be constructed from $e^{2\pi i Z}$ by a ``rotation"
and the latter commutes with the ``rotation'' operator itself [see (\ref{eq:2pi})], the quantity $e^{2\pi i\bm n_0\cdot\bm\Sigma}$ is nothing but $e^{2\pi i Z}$.
Recalling that $Z$ is essentially the number operator $\hat n=\hat a^\dagger\hat a$, we understand that
\begin{equation}
\log(e^{2\pi i\bm n_0\cdot\bm\Sigma})
=\log(e^{i\pi(\hat n+{1\over2})})={i\pi\over2}+i\pi P_\mathrm{odd}
={i\pi\over 2}+i\pi{1-e^{i\pi\hat n}\over2}\equiv{i\pi\over2}\hat q.
\label{eq:hatq}
\end{equation}
The operator $P_{\mathrm{odd}}$ is the projection operator on the odd-number states and commutes with $\bm\Sigma$, for the latter is composed of only quadratic combinations of $\hat a$ and $\hat a^\dagger$.
The operator $\hat q$ defined above has eigenvalues $1$ and $3$ for even- and odd-number states, commutes with any operator spanned by $X$, $Y$, and $Z$, and is necessary to make the equality (\ref{eq:defW}) hold beyond the horizon $x=-1$.
We need a central extension of the algebra by including non-zero $i\alpha I={i\pi\over2}\hat q$ to express the product of two unitaries as a single exponential for ``large'' exponents.

The emergence of a central charge related to the parity operator was to be expected. This is a consequence of the fact that the state space is the direct sum of two irreducible representations of the symplectic Lie algebra sp$(2,\mathbb{R})$,
\begin{equation}
	\mathcal{H} = \mathcal{H}_{\mathrm{even}} \oplus \mathcal{H}_{\mathrm{odd}},
	\qquad 
	\mathcal{H}_{\mathrm{even}}= \biggl\{ |\psi\rangle=\sum_{k\geq 0}  \psi_{2k}|2k\rangle \biggr\},
	\qquad 
	\mathcal{H}_{\mathrm{odd}}= \biggl\{ |\psi\rangle=\sum_{k\geq 0}  \psi_{2k+1}|2k+1\rangle\biggr\},
\end{equation}
with $\mathcal{H}_{\mathrm{even(odd)}}$ containing even-number (odd-number) states. Indeed, the generators~\eqref{eq:genrep} of the Lie algebra representation are quadratic in the annihilation and creation operators, and will take even-number vectors to even-number vectors and odd-number to odd-number.

We have seen that there are two $C$'s existing beyond the critical point $\omega=\pi$.
One of them is just a continuous extension from smaller $\omega<\pi$ to larger $\omega>\pi$ and the second is the same as the former at $\pi$ and differs from it for $\omega>\pi$.
While the former faces a singularity at $a=2\pi i$ ($x=-1$), the latter remains finite at $\tilde a=0$ ($x=-1$).
This can be viewed as a kind of bifurcation: a proper choice, leading to a well-defined $C$ for all values of parameters, is possible at the expense of a discontinuity in the parameter $a$ and the introduction of the new element $\hat q$. The latter (proper branch) brings about $C=\eta Y$ at even larger $\omega=2\pi$, where we have $\bar a=\eta$ and $\xi=0$.
It is stressed that the point $\omega=\pi$ is unique, in the sense that since the shift by $2\pi i$ in $a$ yields only a sign change in $\cosh{a\over2}=x$, both of them are solutions only at $x=0$ corresponding to $\omega=\pi$, which allows us to choose an alternative branch at $\omega=\pi$, keeping the exponent $C$ continuous.   
The above procedure has to be repeated whenever $a$ approaches $\pi i$ (i.e., when $\cos{\omega\over2}=0$) and thus the product of two unitaries can be expressed as a single-exponential function in all parameter regions, provided an appropriate extension of the original algebra (\ref{eq:AMalgebra}) is made.

\subsection{Explicit Expressions of $C$ for $0\le\omega\le2\pi$}
The exponent $C$ explicitly reads 
\begin{align}
C&=Xa\cosh\xi\sin{\omega\over2}+Ya\cosh\xi\cos{\omega\over2}+iZa\sinh\xi,\\
&\sinh{a\over2}\cosh\xi=\sinh{\eta\over2},\\
&\sinh{a\over2}\sinh\xi=\cosh{\eta\over2}\sin{\omega\over2},\\
&\cosh{a\over2}=\cosh{\eta\over2}\cos{\omega\over2}=x,\qquad\text{for $1\le x$ or $0\le\omega\le \omega_1$ [$a: 0\to0$],}
\end{align}
\begin{align}
C&=X\bar a\sinh\tilde\xi\sin{\omega\over2}+Y\bar a\sinh\tilde\xi\cos{\omega\over2}+iZ\bar a\cosh\tilde \xi,\\
&\sinh{i\bar a\over2}\cosh(\tilde\xi-{\pi\over2}i)=\sin{\bar a\over2}\sinh\tilde\xi=\sinh{\eta\over2},\\
&\sin{\bar a\over2}\cosh\tilde\xi=\cosh{\eta\over2}\sin{\omega\over2},\\
&\cos{\bar a\over2}=\cosh{\eta\over2}\cos{\omega\over2},\qquad\text{for $x:1\to0$ or $\omega_1\le\omega\le\pi$ [$\bar a: 0\to\pi$],}
\end{align}
\begin{align}
C&=X\tilde a\sinh\tilde\xi\sin{\omega\over2}+Y\tilde a\sinh\tilde\xi\cos{\omega\over2}+iZ\tilde a\cosh\tilde \xi+{i\pi\over2}\hat q,\\
&\sinh{i\tilde a+2\pi i\over2}\cosh(\tilde\xi-{\pi\over2}i)=-\sin{\tilde a\over2}\sinh\tilde\xi=\sinh{\eta\over2},\\
&-\sin{\tilde a\over2}\cosh\tilde\xi=\cosh{\eta\over2}\sin{\omega\over2},\\
&-\cos{\tilde a\over2}=\cosh{\eta\over2}\cos{\omega\over2},\qquad\text{for $x: 0\to-1$ or $\pi\le\omega\le\omega_{-1}<2\pi$ [$\tilde a: -\pi\to0$],}
\end{align}
\begin{align}
C&=-Xa\cosh\tilde\xi\sin{\omega\over2}-Ya\cosh\tilde\xi\cos{\omega\over2}-iZa\sinh\tilde \xi+{i\pi\over2}\hat q,
\\
&\sinh{a+2\pi i\over2}\cosh(\tilde\xi-\pi i)=\sinh{a\over2}\cosh\tilde\xi=\sinh{\eta\over2},\\
&\sinh{a\over2}\sinh\tilde\xi=\cosh{\eta\over2}\sin{\omega\over2},\\
&-\cosh{a\over2}=\cosh{\eta\over2}\cos{\omega\over2},\qquad\text{for $x\le-1$ or $\omega_{-1}\le\omega\le2\pi$ [$a: 0\to\eta$].}
\end{align}

\subsection{Explicit Expressions of the Coefficients}
We summarize here the explicit expressions of the coefficients that appear in the definitions of $h$ in the main text [Eqs.\ (9) and (14)].
Let
\begin{equation}
e^{i\omega\hat{h}_H}e^{i\eta\hat{h}_-}=e^{i\hat{h}},
\end{equation}
\begin{equation}
\hat{h}_H=\frac{1}{4}(\hat{a}^\dag\hat{a}+\hat{a}\hat{a}^\dag),\quad
\hat{h}_+=\frac{1}{4}(\hat{a}^{\dag2}+\hat{a}^2),\quad
\hat{h}_-=\frac{1}{4i}(\hat{a}^{\dag2}-\hat{a}^2).
\end{equation}
At variance with the main text, we shall use here indices $1$ and $2$ rather than primes: therefore the variables 
$\alpha_1$, $\beta_1$, $\gamma_1$ used below become $\alpha$, $\beta$, $\gamma$ in the main text, and the 
variables 
$\alpha_2$, $\beta_2$, $\gamma_2$ used below become $\alpha'$, $\beta'$, $\gamma'$ in the main text.

\begin{figure}[b]
\centering
\begin{tabular}{lll}
(a)&(b)&(c)\\
\includegraphics[height=0.3\textwidth]{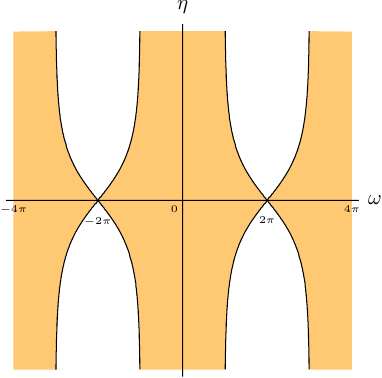}&
\includegraphics[height=0.3\textwidth]{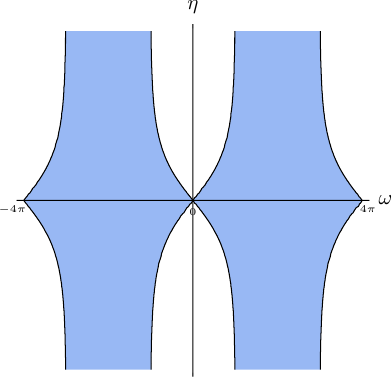}&
\includegraphics[height=0.3\textwidth]{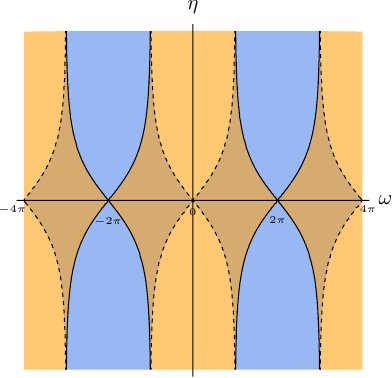}
\end{tabular}
\caption{Ranges of validity for the different expressions of the coefficients $\alpha, \beta, \gamma$.
(a) Main branch. (b) Second branch, inside the horizon. (c) Overlap.
}
\label{branches}
\end{figure}

%\bigskip
\begin{itemize}
\item
First branch without $\hat{q}$ (for $x=\cosh\frac{\eta}{2}\cos\frac{\omega}{2}>-1$):
\begin{equation}
\hat{h}=\hat{h}_1=\alpha_1(\omega,\eta)\hat{h}_++\beta_1(\omega,\eta)\hat{h}_-+\gamma_1(\omega,\eta)\hat{h}_H,
\end{equation}
with
\begin{equation}
\begin{cases}
\medskip
\displaystyle
\alpha_1
=a_1\cosh\xi_1\sin\frac{\omega}{2}
=\tilde{a}_1\sinh\tilde{\xi}_1\sin\frac{\omega}{2},
\\
\medskip
\displaystyle
\beta_1
=a_1\cosh\xi_1\cos\frac{\omega}{2}
=\tilde{a}_1\sinh\tilde{\xi}_1\cos\frac{\omega}{2},
\\
\displaystyle
\gamma_1
=a_1\sinh\xi_1
=\tilde{a}_1\cosh\tilde{\xi}_1,
\end{cases}
\end{equation}
\begin{equation}
\begin{cases}
\medskip
\displaystyle
a_1=2\epsilon(\eta)\cosh^{-1}\!\left(
\cosh\frac{\eta}{2}\cos\frac{\omega}{2}
\right),
\\
\displaystyle
\xi_1=\tanh^{-1}\!\left(
\coth\frac{\eta}{2}\sin\frac{\omega}{2}
\right),
\end{cases}
\quad
\begin{cases}
\medskip
\displaystyle
\tilde{a}_1=2\epsilon\!\left(\sin\frac{\omega}{2}\right)
\cos^{-1}\!\left(
\cosh\frac{\eta}{2}\cos\frac{\omega}{2}
\right),
\\
\displaystyle
\tilde{\xi}_1=\coth^{-1}\!\left(
\coth\frac{\eta}{2}\sin\frac{\omega}{2}
\right),
\end{cases}
\end{equation}
where $\epsilon(s)=s/|s|$.
This solution is valid in the orange region in Fig.\ \ref{branches}(a). We call it the main branch.

\item
Second branch with central extension $\hat{q}$ (for $x=\cosh\frac{\eta}{2}\cos\frac{\omega}{2}<1$):
\begin{equation}
\hat{h}=\hat{h}_2=\alpha_2(\omega,\eta)\hat{h}_++\beta_2(\omega,\eta)\hat{h}_-+\gamma_2(\omega,\eta)\hat{h}_H+\frac{\pi}{2}\hat{q},
\qquad
\hat{q}=2-(-1)^{\hat{n}},
\end{equation}
with
\begin{equation}
\begin{cases}
\medskip
\displaystyle
\alpha_2
=a_2\cosh\xi_2\sin\frac{\omega}{2}
=\tilde{a}_2\sinh\tilde{\xi}_2\sin\frac{\omega}{2},
\\
\medskip
\displaystyle
\beta_2
=a_2\cosh\xi_2\cos\frac{\omega}{2}
=\tilde{a}_2\sinh\tilde{\xi}_2\cos\frac{\omega}{2},
\\
\displaystyle
\gamma_2
=a_2\sinh\xi_2
=\tilde{a}_2\cosh\tilde{\xi}_2,
\end{cases}
\end{equation}
\begin{equation}
\begin{cases}
\medskip
\displaystyle
a_2=-2\epsilon(\eta)
\cosh^{-1}\!\left(
-\cosh\frac{\eta}{2}\cos\frac{\omega}{2}
\right),
\\
\displaystyle
\xi_2=\tanh^{-1}\!\left(
\coth\frac{\eta}{2}\sin\frac{\omega}{2}
\right),
\end{cases}
\quad
\begin{cases}
\medskip
\displaystyle
\tilde{a}_2=-2\epsilon\!\left(\sin\frac{\omega}{2}\right)\cos^{-1}\!\left(
-\cosh\frac{\eta}{2}\cos\frac{\omega}{2}
\right),
\\
\displaystyle
\tilde{\xi}_2=\coth^{-1}\!\left(
\coth\frac{\eta}{2}\sin\frac{\omega}{2}
\right).
\end{cases}
\end{equation}
\end{itemize}
This solution is valid in the blue region in Fig.\ \ref{branches}(b).

The overlap between the regions of validity of the two solutions is shown in Fig.\ \ref{branches}(c).
Coefficients $\alpha$, $\beta$, and $\gamma$ are shown in Figs.\ 
\ref{alpha1_math}, \ref{beta1_math}, and \ref{gamma1_math}, respectively.

\begin{figure}[h]
\centering
\begin{tabular}{lll}
(a)&(b)&(c)\\
\includegraphics[height=0.24\textwidth]{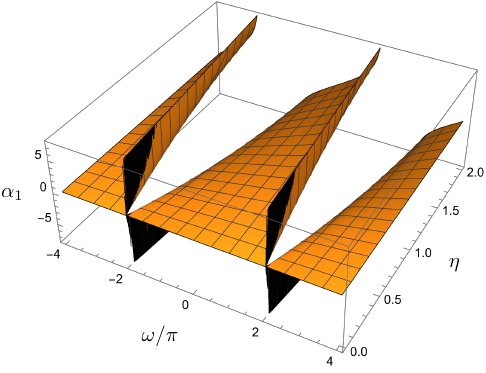}&
\includegraphics[height=0.24\textwidth]{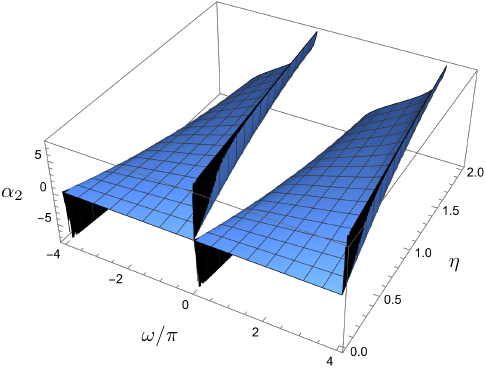}&
\includegraphics[height=0.24\textwidth]{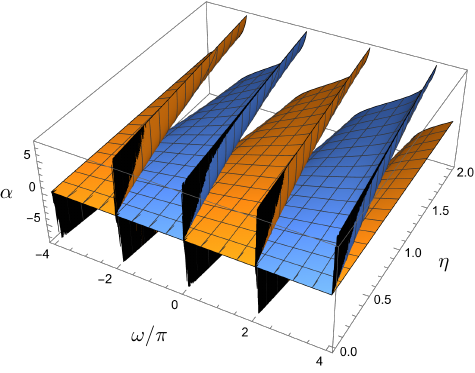}
\end{tabular}
\caption{Coefficient $\alpha$. (a) Main branch: notice the horizon. 
(b) Inside the horizon. (c) Both solutions. The colors are the same as in Fig.\ \ref{branches} of this Supplemental Material and Fig.\ 3 of the main text.
 }
\label{alpha1_math}
\end{figure}
\begin{figure}[h]
\begin{tabular}{lll}
(a)&(b)&(c)\\
\includegraphics[height=0.24\textwidth]{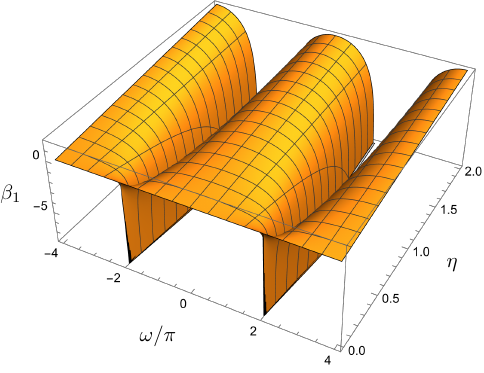}&
\includegraphics[height=0.24\textwidth]{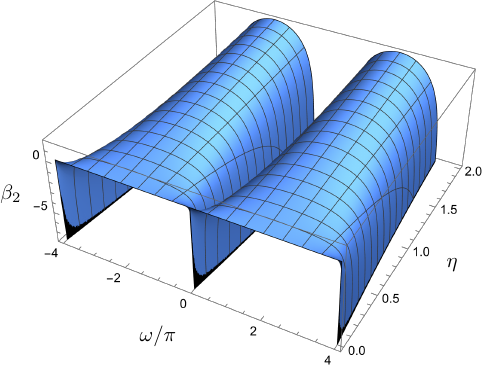}&
\includegraphics[height=0.24\textwidth]{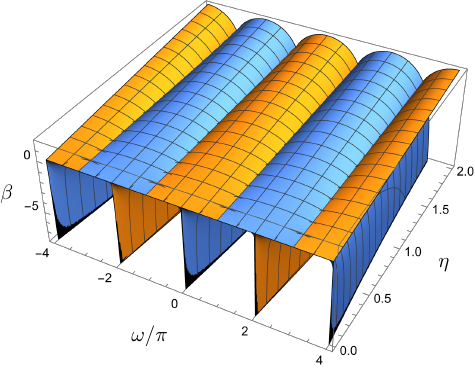}
\end{tabular}
\caption{Coefficient $\beta$. (a) Main branch. 
(b) Inside the horizon. (c) Both solutions. The colors are the same as in Fig.\ \ref{branches} of this Supplemental Material and Fig.\ 3 of the main text.}
\label{beta1_math}
\end{figure}
\begin{figure}[h]
\begin{tabular}{lll}
(a)&(b)&(c)\\
\includegraphics[height=0.24\textwidth]{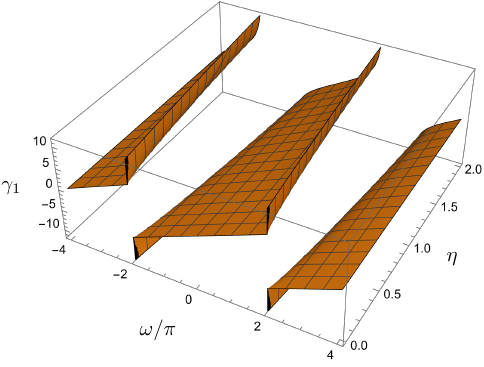}&
\includegraphics[height=0.24\textwidth]{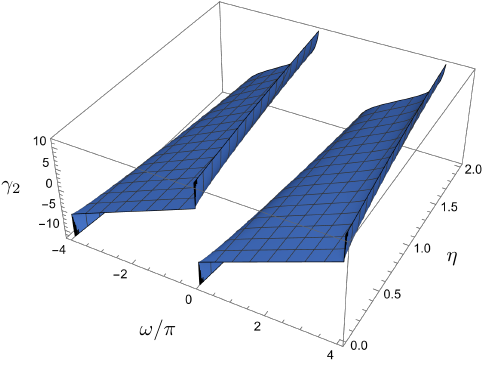}&
\includegraphics[height=0.24\textwidth]{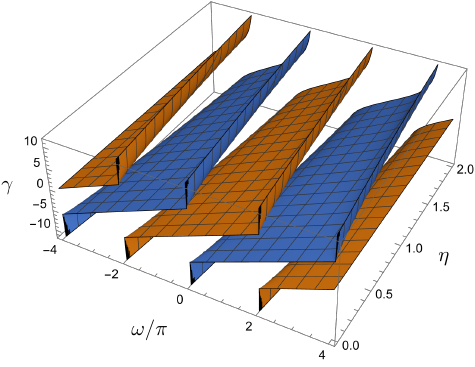}
\end{tabular}
\caption{Coefficient $\gamma$. (a) Main branch. 
(b) Inside the horizon. (c) Both solutions. The colors are the same as in Fig.\ \ref{branches} of this Supplemental Material and Fig.\ 3 of the main text.}
\label{gamma1_math}
\end{figure}

\subsection{Case of Angular Momenta}
When all the operators $X$, $Y$, and $Z$ are Hermitian, they are essentially the angular-momentum operators because of the commutation relations they satisfy.
We denote them as $\bm J$ and look for an anti-Hermitian $C$ that satisfies
\begin{equation}
e^{i\omega tJ_z}e^{i\eta tJ_y}=e^{C(t)}.
\end{equation}
It is now parametrized as
\begin{equation}
C(t)=ia(t)\bm n(t)\cdot\bm J,
\end{equation}
where $\bm n=\cos\xi(\bm e_1\cos\varphi+\bm e_2\sin\varphi)+\bm e_3\sin\xi$ and all quantities $\omega$, $\eta,a(t)$, $\xi(t)$, and $\varphi(t)$ are real parameters and real functions of $t$.
The solution can be readily obtained from the previous one,  just by replacing $\omega\to\omega$, $\eta\to i\eta$, $a\to ia$, $\xi\to-i\xi$, $\varphi\to\varphi$.
We obtain $\varphi={\pi\over2}-{\omega t\over2}$ and
\begingroup
\allowdisplaybreaks
\begin{align}
\sin{a\over2}\cos\xi&=\sin{\eta t\over2},\\
\sin{a\over2}\sin\xi&=\cos{\eta t\over2}\sin{\omega t\over2},\\
\cos{a\over2}&=\cos{\eta t\over2}\cos{\omega t\over2}.
\end{align}
\endgroup
It is evident that these equations allow real-valued solutions for any values of $\omega$ and $\eta$ ($t=1$), and no singular behavior occurs.
\end{widetext}
\end{document}